\documentstyle[aps,prd,epsfig,preprint]{revtex}
\def\fun#1#2{\lower3.6pt
\vbox{\baselineskip0pt\lineskip.9pt
\ialign{$\mathsurround=0pt#1\hfill##\hfil$
\crcr#2\crcr\sim\crcr}}}
\begin{document}
\vspace{0.5in}
\title{\vskip-2.5truecm{\hfill \baselineskip 14pt 
{\hfill {{\small \hfill UT-STPD-7/99 }}}\vskip .1truecm} 
\vspace{1.0cm}
\vskip 0.1truecm {\bf Supersymmetric Cold Dark Matter 
with Yukawa Unification}}
\vspace{1cm}
\author{{M. E. G\'omez}\thanks{mgomez@cc.uoi.gr},
{G. Lazarides}\thanks{lazaride@eng.auth.gr} 
{and C. Pallis}\thanks{kpallis@gen.auth.gr}} 
\vspace{1.0cm}
\address{{\it Physics Division, School of Technology, 
Aristotle University of Thessaloniki,\\ 
Thessaloniki GR 540 06, Greece.}}
\maketitle

\vspace{2cm}

\begin{abstract}
\baselineskip 12pt

\par
The cosmological relic density of the lightest 
supersymmetric particle of the minimal supersymmetric 
standard model is calculated under the assumption of 
gauge and Yukawa coupling unification. We employ radiative 
electroweak breaking with universal boundary conditions 
from gravity-mediated supersymmetry breaking. Coannihilation 
of the lightest supersymmetric particle, which turns out to 
be an almost pure bino, with the next-to-lightest 
supersymmetric particle (the lightest stau) is crucial 
for reducing its relic density to an acceptable level. 
Agreement with the mixed or the pure cold (in the presence 
of a nonzero cosmological constant) dark matter scenarios 
for large scale structure formation in the universe requires 
that the lightest stau mass is about $2-8\%$ larger than 
the bino mass, which can be as low as 222 GeV. The smallest 
allowed value of the lightest stau mass turns out to be about 
232 GeV.
\end{abstract}

\thispagestyle{empty}
\newpage
\pagestyle{plain}
\setcounter{page}{1}
\baselineskip 20pt

\section{Introduction}
\label{sec:intro}

\par
It is by now clear \cite{mdm} that, in a universe with 
zero cosmological constant, a combination of cold plus 
hot dark matter is needed for fitting the data on cosmic 
microwave background (CMB) anisotropies and large scale 
structure \cite{structure} in the universe, especially 
for essentially flat spectrum of the primordial density 
fluctuations. The energy density $\rho$ of the universe 
is taken equal to its critical value $\rho_{c}$ ($\Omega
\equiv\rho/\rho_{c}=1$), as suggested by inflationary 
cosmology, and assumed to consist solely of matter 
($\Omega_{m}=1$). About 10$\%$ of matter is baryonic 
($\Omega_{B}\approx 0.1$), while the rest (dark matter) 
contains a hot component with density equal to about 20$\%$ 
of the critical density ($\Omega_{HDM} \approx 0.2$) and 
a cold one with $\Omega_{CDM}\approx 0.7$. The present 
value of the Hubble parameter in units of $100~\rm{km}
~\rm{sec}^{-1}~\rm{Mpc}^{-1}$ is taken to be 
$h\approx 0.5$. Hot dark matter may consist of light 
neutrinos. This is compatible with the atmospheric 
\cite{superk} and solar neutrino oscillations, within a 
three neutrino scheme, only if light neutrino masses are 
almost degenerate. A consistent supersymmetric inflationary 
model with degenerate light neutrino masses providing the 
hot dark matter in the universe has been constructed in 
Ref.\cite{deg}. Cold dark matter, in the case of vanishing 
cosmological constant, must satisfy the relation 
$\Omega_{CDM}~h^{2}\approx 0.175$.

\par 
Recent observational developments, however, seem to hint 
towards an alternative picture for the composition of the 
energy density of the universe with a nonvanishing 
contribution from something like a cosmological constant. 
Measurements \cite{barfrac} of the cluster baryon fraction 
combined with the low deuterium abundance constraint 
\cite{deuterium} on the baryon asymmetry of the universe, 
$\Omega_{B}~h^{2}\approx 0.02$, suggest that the matter 
density is around 35$\%$ of the critical density of the 
universe ($\Omega_{m}\approx 0.35$). Also, recent 
observations \cite{lambda} favor the existence of a 
cosmological constant, whose contribution to the energy 
density can be as large as 65$\%$ of the critical 
density ($\Omega_{\Lambda}\approx 0.65$), driving 
the total energy density close to its critical value 
as required by inflation. The assumption that dark 
matter contains only a cold component leads then to a 
`good' fit \cite{lambdafit} of the CMB radiation and 
both the large scale structure and age of the universe 
data. Higher values of the Hubble constant ($h \approx 
0.65$) are, however, required and, thus, $\Omega_{CDM}
\approx 0.3$. Moreover, the possibility of 
improving this fit by adding light neutrinos as hot dark 
matter appears \cite{primack} to be rather limited. We 
can, thus, assume hierarchical neutrino masses in this 
case. A consistent supersymmetric picture leading 
`naturally' to hybrid inflation and employing hierarchical 
neutrino masses has been presented in Ref.\cite{hier}. 
In the presence of a nonvanishing cosmological constant, 
cold dark matter must satisfy $\Omega_{CDM}~h^{2}
\approx 0.125$.

\par 
Both these cosmological models with zero/nonzero 
cosmological constant, which provide the best fits to all 
the available data, are equally plausible alternatives 
for the composition of the energy density of the universe. 
Thus, taking into account the observational uncertainties, 
we will restrict $\Omega_{CDM}~h^{2}$ in the range 
$0.09-0.22$.

\par
The lightest supersymmetric particle (LSP) of the minimal 
supersymmetric standard model (MSSM) is one of the most 
promising candidates for cold dark matter \cite{gold,ehnos}. 
This is normally the lightest neutralino and its 
stability is guaranteed by the presence of a discrete 
$Z_{2}$ matter parity, which implies that supersymmetric 
particles can disappear only by annihilating in pairs. The 
cosmological relic density of the lightest neutralino can 
be reliably computed, for various values of the parameters 
of MSSM, under the assumptions of gauge coupling unification 
and radiative electroweak breaking with universal boundary 
conditions from gravity-mediated supersymmetry breaking 
(see e.g., Refs.\cite{drees,ellis,lahanas}). Coannihilation 
\cite{coan} of the LSP with the next-to-lightest 
supersymmetric particle (NLSP) turns out to be crucial in 
many cases \cite{drees,ellis,gondolo}.

\par
The purpose of this paper is to estimate the lightest 
neutralino relic density in a specific MSSM framework 
\cite{als} of the above variety, where the three Yukawa 
couplings of the third family of quarks and leptons unify 
`asymptotically' (i.e., at the grand unified theory (GUT) 
scale $M_{GUT}\sim 10^{16}~{\rm{GeV}}$). 
This can arise by embedding MSSM in a supersymmetric GUT 
based on a gauge group such as $SO(10)$ or $E_{6}$, 
where all the particles of one family belong to a single 
representation. It is then obvious that requiring the 
masses of the third family fermions to arise primarily 
from their unique Yukawa coupling to a particular 
superfield representation (say a 10-plet of $SO(10)$) 
predominantly containing the electroweak 
higgses guarantees the desired Yukawa coupling unification. 
This scheme predicts large $\tan\beta\approx m_{t}/m_{b}$, 
as well as the successful `asymptotic' mass relation 
$m_{\tau}=m_{b}$. The supersymmetric particle spectrum, 
top quark mass and higgs scalar masses in this model have 
been studied in Refs.\cite{spartop,copw,anant}. The top 
quark mass is `naturally' restricted to large values 
compatible with the present experimental data and the 
supersymmetric particle masses are predicted relatively 
large. The lightest neutralino is an almost pure bino, 
whereas the NLSP is the lightest stau mass eigenstate.

\par
Coannihilation of the bino with the NLSP turns out to 
be of crucial importance for keeping the bino relic density 
at an acceptably low level. This implies that the lightest
stau must not be much heavier than the bino so that 
coannihilation can be effective. Moreover, increasing the 
lightest stau to bino mass ratio leads to a larger bino mass 
which further enhances its relic density. Lightest stau masses 
of about $2-8\%$ larger than the bino mass are required 
for obtaining $\Omega_{CDM}~h^{2}$ in the range $0.09-0.22$.
It is interesting to note that, for smaller `relative' mass 
gaps between the lightest stau and the bino, $\Omega_{CDM}
~h^{2}$ rapidly decreases and becomes unacceptably small.
The values of this mass gap which we find here 
combined with the fact that the bino mass turns out to be 
greater than about 222 GeV make the lightest stau a 
phenomenologically interesting charged sparticle 
with mass which can be as low as $\approx$ 232 GeV. 
Our analysis provides quite strong restrictions on the 
sparticle spectrum of MSSM with Yukawa coupling unification.

\par
In Sec.\ref{sec:model}, the MSSM with Yukawa coupling 
unification is introduced and its parameters and sparticle 
spectrum are constrained. In Sec.\ref{sec:relic}, the 
relic LSP (lightest neutralino) density is calculated by 
taking into account its coannihilation with the NLSP 
(lightest stau). In particular, the bino annihilation 
cross section is estimated in Sec.\ref{subsec:annih}, 
whereas Sec.\ref{subsec:coan} is devoted to the 
evaluation of the relevant coannihilation cross sections. 
Our results on $\Omega_{LSP}~h^2$ are presented and 
their consequences are discussed in 
Sec.\ref{subsec:result}. Finally, our conclusions are 
summarized in Sec.\ref{sec:conclusion}.

\section{MSSM with Yukawa Unification}
\label{sec:model}

\par
We consider the MSSM embedded in some general supersymmetric 
GUT based on a gauge group such as $SO(10)$ or $E_{6}$ 
(where all the particles of one family belong to a single 
representation) with the additional requirement that the top, 
bottom and tau Yukawa couplings unify \cite{als} at the 
GUT scale $M_{GUT}$. This requirement is easily guaranteed
by ensuring that the masses of the third family fermions 
arise primarily from their unique Yukawa coupling to a single 
superfield representation which predominantly contains the 
electroweak higgses. We further assume that the GUT gauge 
symmetry breaking occurs in one step. Ignoring the Yukawa 
couplings of the first and second generation, the effective 
superpotential below $M_{GUT}$ is 
\begin{equation}
W=\epsilon_{ij}(-h_t H_2^i Q_3^j t^c+ h_b H_1^i Q_3^j b^c 
+ h_\tau H_1^i L_3^j \tau^c + \mu H_1^i H_2^j)~,
\label{super}
\end{equation}
where $Q_3=(t,b)$ and $L_3=(\nu_{\tau},\tau)$ are 
the quark and lepton $SU(2)_{L}$ doublet left handed 
superfields of the third generation and $t^c$, $b^c$ 
and $\tau^c$ the corresponding $SU(2)_{L}$ singlets. 
Also, $H_1$, $H_2$ are the electroweak higgs superfields 
and $\epsilon_{12}=+1$. The gravity-mediated soft 
supersymmetry breaking terms in the scalar potential are 
given by 
$$    
V_{soft}= \sum_{a,b} m_{ab}^{2}\phi^{*}_a \phi_b+ 
$$
\begin {equation}   
\left(\epsilon_{ij}(-A_t h_t  H_2^i \tilde Q_3^j \tilde t^c
+A_b h_b H_1^i \tilde Q_3^j\tilde b^c 
+A_\tau h_\tau H_1^i \tilde L_3^j \tilde\tau^c
+ B\mu H_1^iH_2^j)+ h.c.\right)~,
\label{vsoft}
\end{equation}
where the $\phi_a$ 's are the (complex) scalar fields and 
tildes denote superpartners. The gaugino mass terms in the 
Lagrangian are 
\begin{equation}
\frac{1}{2}(M_1 \tilde B\tilde B + 
M_2 \sum_{r=1}^{3} \tilde W_r \tilde W_r +  
M_3 \sum_{a=1}^{8}\tilde g_a\tilde g_a+h.c.)~,
\label{gaugino}
\end{equation}
where $\tilde B$, $\tilde W_r$ and $\tilde g_a$ are 
the bino, winos and gluinos respectively. `Asymptotic' 
Yukawa coupling unification implies
\begin{equation}
h_t(M_{GUT})=h_b(M_{GUT})=h_{\tau}(M_{GUT})\equiv h_0~.
\label{yukawa}
\end{equation}
Based on $N=1$ supergravity, we take universal soft 
supersymmetry breaking terms at $M_{GUT}$, i.e., a 
common mass for the scalar fields $m_0$, a common 
trilinear scalar coupling $A_0$ and $B_0=A_0-m_0$. 
Also, a common gaugino mass $M_{1/2}$ is assumed at 
$M_{GUT}$.

\par
Our effective theory below $M_{GUT}$ depends on the 
parameters ($\mu_0=\mu(M_{GUT})$)
\[
m_0,\ M_{1/2},\ A_0,\ \mu_0,\ \alpha_G,\ M_{GUT},
\ h_{0},\ \tan\beta~.  
\]
The quantities $\alpha_G=g_{G}^{2}/4\pi$ ($g_{G}$ being 
the GUT gauge coupling constant) and $M_{GUT}$ are evaluated 
consistently with the experimental values of $\alpha_{em},\ 
\alpha_s$ and $\sin^2\theta_W$ at $m_Z$. We integrate 
numerically the renormalization group equations (RGEs) 
for the MSSM at two loops in the gauge and Yukawa couplings 
from $M_{GUT}$ down to a common supersymmetry threshold 
$M_S\sim 1~{\rm{TeV}}$. From this energy to $m_Z$, the 
RGEs of the nonsupersymmetric standard model are used. The 
set of RGEs needed for our computation can be found in many 
references (see, for example, Ref.\cite{rge}). We take 
$\alpha_s(m_Z)=0.12 \pm 0.001$ which, as it turns out, 
leads to gauge coupling unification at $M_{GUT}$ with an 
accuracy better than 0.1$\%$. This allows us to assume an 
exact unification once the appropriate supersymmetric 
particle thresholds are taken into account. Our integration 
procedure relies on iterative runs of the RGEs from $M_{GUT}$ 
to low energies and back, for every set of values of the input 
parameters, until agreement with the experimental data is 
achieved. The value of $\tan\beta$ at $M_S$ is estimated  
using the experimental input $m_\tau(m_{\tau})=1.777~
{\rm{ GeV}}$ and $M_S$ is fixed to be $1~{\rm{TeV}}$ 
throughout our calculation. Assuming radiative electroweak 
symmetry breaking, we can express the values of the parameters 
$\mu$ (up to its sign) and $B$ at $M_S$ in terms of the 
other input parameters by means of the appropriate conditions
\begin{equation}
\mu^2=\frac{m^2_{H_1}-m^2_{H_2}\tan^2{\beta}}
{\tan^2{\beta}-1}- \frac{1}{2} m^2_Z \ , \ \sin 2\beta
=-\frac{2 B \mu}{m_{H_1}^2+m_{H_2}^2+2 \mu^2}~,
\label{mu}
\end{equation}
where $m_{H_1}$, $m_{H_2}$ are the soft 
supersymmetry breaking scalar higgs masses. Here, following 
Ref.\cite{grz}, we used the tree-level renormalization 
group improved scalar potential minimized at a scale 
comparable to the mass of the stop quark. This is adequate 
for our purposes since, as we find, the corrections 
to $\mu$ from the full one-loop effective potential in 
Ref.\cite{pierce} are negligible. The sign of $\mu$ is 
taken to be negative (with our conventions), which leads 
to acceptable predictions for $b\rightarrow s\gamma$ in 
models with large $\tan\beta$ \cite{bbct}.

\par
The common value of the third generation Yukawa coupling 
at $M_{GUT}$ is found by fixing the top quark mass at 
the center of its experimental range, $m_t(m_t)=166~
{\rm{GeV}}$. The value obtained for $m_b(m_Z)$ after 
including supersymmetric corrections is somewhat higher 
than the experimental limit \cite{mb}. We are left with 
$m_0,\ M_{1/2}$ and $A_0$ as free input parameters. 
Our results, as it turns out, depend very little on the 
exact value of $A_0$ which is, thus, fixed to zero in 
our calculation. The values of $m_0$ and 
$M_{1/2}$ are found as functions of the tree-level mass 
$m_A$ of the CP-odd higgs scalar $A$, for each `relative' 
mass splitting between the NLSP (lightest stau) and the 
LSP (almost a pure bino), as we will explain later. The 
value of $m_A$ is evaluate at $M_S$ which is comparable 
with $\sqrt{m_{\tilde{t}}m_{\tilde t^c}}$ \cite{bcdpt}. 
Although the full one-loop corrections to $m_A$ (from 
Ref.\cite{pierce}) are not totally negligible, we will 
ignore them here since their effect on the LSP relic 
density is small.

\par 
The LSP is the lightest neutralino $\tilde{\chi}$. The mass 
matrix for the four neutralinos is  
\begin{eqnarray}
\left(\matrix{ 
M_1 & 0 & -m_Z s_W\cos\beta & m_Z s_W\sin\beta
\cr
0 & M_2 & m_Z c_W\cos\beta & -m_Z c_W\sin\beta
\cr
-m_Z s_W\cos\beta & m_Z c_W\cos\beta & 0 & \mu
\cr
m_Z s_W\sin\beta & -m_Z c_W\sin\beta & \mu & 0
\cr}\right)~,
\label{neutralino}
\end{eqnarray}
in the $(-i\tilde B, -i\tilde W_3, \tilde H_1, 
\tilde H_2)$ basis. Here $s_W=\sin \theta_W$, 
$c_W=\cos \theta_W $, and $M_1$, $M_2$ are the mass 
parameters of $\tilde B$, $\tilde W_3$ in 
Eq.(\ref{gaugino}). For the values of $\mu$ 
obtained from the radiative electroweak breaking conditions 
here ($\mu/M_{1/2}\approx 1.2$), the lightest neutralino 
turns out to be a bino, $\tilde B$, with purity $>98\%$.

\par
Large $b$ and $\tau$ Yukawa couplings cause soft 
supersymmetry breaking masses of the third generation squarks 
and sleptons to run (at low energies) to lower physical values 
than the corresponding masses of the first and second 
generation. Furthermore, the large values of $\tan\beta$ 
implied by the unification of the third generation Yukawa 
couplings lead to large off-diagonal mixings in the sbottom and 
stau mass-squared matrices. These effects make the physical mass 
of the lightest stau significantly lower than the masses of the 
other squarks and sleptons (see below). The NLSP is, thus, the 
lightest stau mass eigenstate $\tilde\tau_2$ and its mass is 
obtained by diagonalizing the stau mass-squared matrix
\begin{eqnarray}
\left(\matrix{
m_{\tau}^2+m_{\tilde\tau_L}^2+m_Z^2 (-1/2+s_W^2)\cos 2\beta 
& m_{\tau}(A_{\tau}+\mu\tan\beta)\cr 
m_{\tau}(A_{\tau}+\mu\tan\beta)
& m_{\tau}^2+m_{\tilde\tau_R}^2-m_Z^2 s_W^2\cos 2\beta\cr}
\right)~,
\label{stau}
\end{eqnarray}
in the gauge basis ($\tilde\tau_L,\ \tilde\tau_R$). Here, 
$m_{\tilde\tau_{L(R)}}$ is the soft supersymmetry breaking 
mass of $\tilde\tau_{L(R)}$ and $m_{\tau}$ the tau lepton 
mass. The stau mass eigenstates are
\begin{eqnarray}
\left( \matrix{ 
\tilde\tau_1  
\cr 
\tilde\tau_2 
\cr} 
\right) 
& = &
\left( \matrix{ 
c_\theta & s_\theta
\cr
-s_\theta & c_\theta
\cr} 
\right) 
\left( \matrix{
\tilde\tau_L  
\cr 
\tilde\tau_R 
\cr} 
\right)~,
\label{eigen}   
\end{eqnarray}
where $s_\theta=\sin \theta$, 
$c_\theta=\cos \theta$, with $\theta$ being the 
$\tilde\tau_L-\tilde\tau_R$ mixing angle. Another 
effect of the large values of the $b$ and $\tau$ Yukawa 
couplings is the reduction of the mass of the CP-odd 
higgs boson $m_A$ and, consequently, the other higgs 
boson masses to smaller values. 

\par
The authors of Ref.\cite{anant} found that, for every 
value of $m_A$ and a fixed value of $m_t(m_t)$, there 
is a pair of minimal values of $m_0$ and $M_{1/2}$ 
where the masses of the LSP and $\tilde\tau_2$ 
are equal. This is understood from 
the dependence of $m_A$ on $m_0$ and $M_{1/2}$ given in 
Ref.\cite{copw}:
\begin{equation}
m^2_A = \alpha M^2_{1/2}-\beta m^2_0-{\rm{const.}}~,
\label{mAc}
\end{equation} 
where all the coefficients are positive and $\alpha$ and 
$\beta$, which depend only on $m_t(m_t)$, are $\sim 0.1$
(the constant turns out to be numerically close to $m_Z^2$). 
Equating the masses of the LSP and 
$\tilde\tau_2$ is equivalent to relating $m_0$ and 
$M_{1/2}$. Then, for every $m_A$, a pair of values of 
$m_0$ and $M_{1/2}$ is determined. Note that 
Eq.(\ref{mAc}) implies the existence of an 
upper bound on $m_A$ since $m^2_A < \alpha M^2_{1/2}~$. 
We set here an upper limit on $M_{1/2}$ equal to 
$800~{\rm{GeV}}$, which keeps the sparticle 
masses below about $2~{\rm{TeV}}$ consistently with 
our choice for $M_S$ (=1 TeV). This limit constrains 
$m_A$ to be smaller than $\approx 220~{\rm{GeV}}$. On 
the other hand, the experimental searches for the lightest 
CP-even neutral higgs boson $h$ with mass $m_{h}$ set a 
lower limit on $m_A$. Taking into account radiative 
corrections \cite{dnr,erz} in calculating $m_{h}$, we 
found that this lower limit on $m_A$ is about 
$95~{\rm{GeV}}$. The highest values of $m_{h}$, which are 
obtained as $m_A$ increases to its upper limit, lie between 
125 and 130 GeV.

\par
\begin{figure}
\epsfig{figure=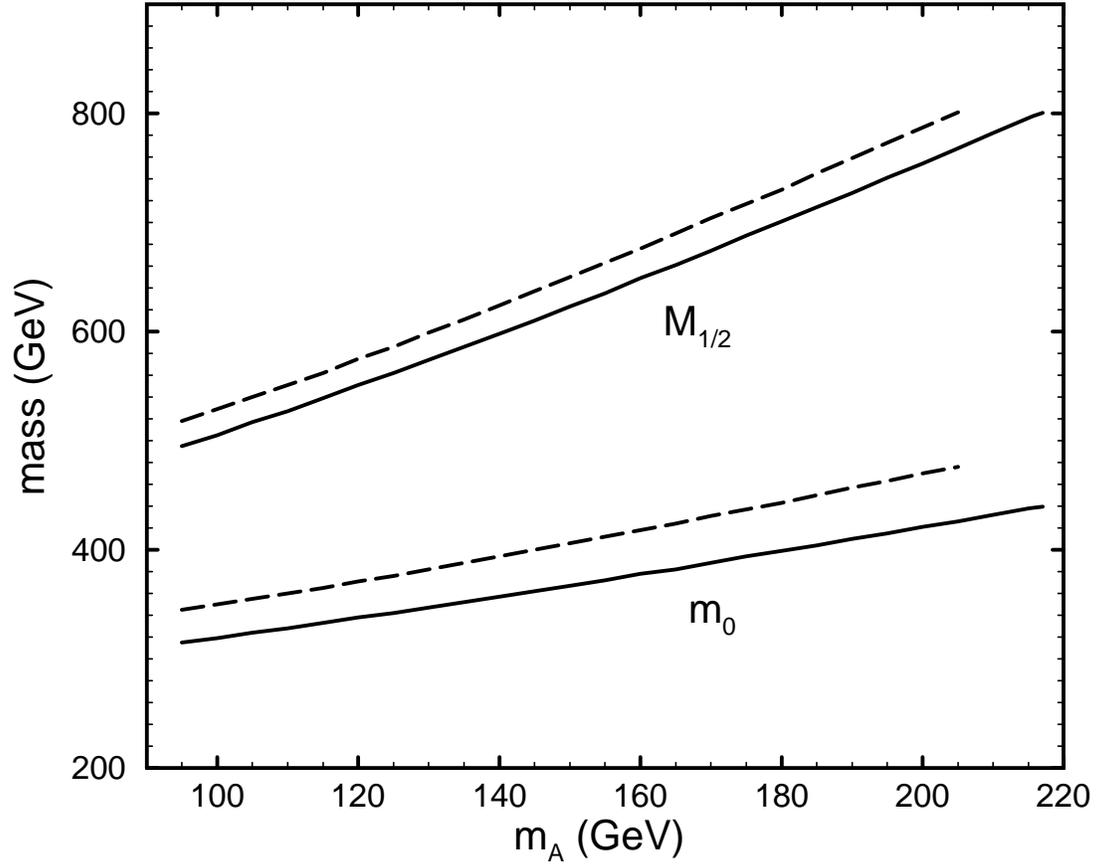,height=4.62in,angle=0}
\medskip
\caption{The mass parameters $m_0$ and $M_{1/2}$ as 
functions of $m_A$ for $\Delta_{\tilde\tau_2}=0.02$ 
(solid lines) and 0.08 (dashed lines).
\label{masses}}
\end{figure}

 
Following the procedure of Ref.\cite{anant}, one can 
determine $m_0$ and $M_{1/2}$ not only for equal 
masses of the LSP and NLSP but for any relation 
between these masses. We fix $m_t(m_t)=166~{\rm{GeV}}$ 
($\tan\beta\approx 52.9$). For every $m_A$ and a 
given `relative' mass splitting $\Delta_{\tilde\tau_2}
=(m_{\tilde\tau_2}-m_{\tilde\chi})/m_{\tilde\chi}$ 
between the NLSP and LSP, we find $m_0$ and $M_{1/2}$. 
They are depicted in Fig.\ref{masses} as functions of 
$m_A$ for $\Delta_{\tilde\tau_2}$=0.02 and 
0.08 (see Sec.\ref{subsec:result}). We observe 
that, for fixed $m_A$, $M_{1/2}$ increases with  
$\Delta_{\tilde\tau_2}$. Thus, $m_0$ and the sparticle 
masses increase too with $\Delta_{\tilde\tau_2}$ (see 
Eq.(\ref{mAc})). Also, for fixed $M_{1/2}$, $m_A$ 
is a decreasing function of $\Delta_{\tilde\tau_2}$. 
As a consequence, the upper bound on $m_A$ 
(corresponding to $M_{1/2}=800~{\rm{GeV}}$) gets 
reduced as $\Delta_{\tilde\tau_2}$ increases. This 
is  why the curves in Fig.\ref{masses} which correspond 
to higher $\Delta_{\tilde\tau_2}$ 's terminate at 
smaller $m_A$ 's. As we will see, the cosmological 
bounds on $\Omega_{LSP}~h^2$ will constrain 
$\Delta_{\tilde\tau_2}$. The relevant part of the 
sparticle spectrum as a function of $m_A$, for 
$\Delta_{\tilde\tau_2}$=0.047, is shown in 
Fig.\ref{spectrum}. The LSP mass, for 
$\Delta_{\tilde\tau_2}$=0.02, is also included. 

\section{LSP Relic Density}
\label{sec:relic}

We now turn to the calculation of the cosmological relic 
density of the lightest neutralino $\tilde\chi$ (almost 
pure $\tilde B$) in MSSM with Yukawa unification. 
As mentioned in Sec.\ref{sec:intro}, 
$\Omega_{\tilde\chi}~h^2$ increases to unacceptably 
high values as $m_{\tilde\chi}$ becomes larger. 
Low values of $m_{\tilde\chi}$ are obtained when 
the NLSP ($\tilde\tau_2$) is almost degenerate with 
$\tilde\chi$. Under these circumstances, 
coannihilation of $\tilde\chi$ with $\tilde\tau_2$ and 
$\tilde\tau_2^\ast$ is of crucial importance 
reducing further the $\tilde\chi$ relic density by a 
significant amount. The important role of coannihilation of 
the LSP with sparticles carrying masses close to its mass in 
the calculation of the LSP relic density has been pointed 
out by many authors (see e.g., Refs.\cite{drees,ellis,coan}). 
Here, we will use the method described by Griest and Seckel 
\cite{coan}. Note that our analysis can be readily applied 
to any MSSM scheme where the LSP and NLSP are the bino and 
stau respectively.

\par
The relevant quantity, in our case, is the total number 
density 
\begin{equation}
n= n_{\tilde\chi} + n_{\tilde\tau_2} 
+ n_{\tilde\tau_2^\ast}~,
\label{n}
\end{equation}
since the $\tilde\tau_2$ 's and $\tilde\tau_2^\ast$ 's 
decay into $\tilde\chi$ 's after freeze-out. At cosmic 
temperatures relevant for freeze-out, the scattering rates 
of these (nonrelativistic) sparticles off particles in 
the thermal bath are much faster than their annihilation 
rates since the (relativistic) particles in the bath are 
considerably more abundant. Consequently, the number 
densities $n_i$ ($i=\tilde\chi$, $\tilde\tau_2$, 
$\tilde\tau_2^\ast$) are proportional to their 
equilibrium values $n_i^{eq}$ to a good approximation, 
i.e., $n_i/n\approx n_i^{eq}/n^{eq}\equiv r_i$. The 
Boltzmann equation (see e.g., Ref.\cite{kt}) is then 
written as
\begin{equation}
\frac{dn}{dt}=-3Hn-\langle \sigma_{eff} v \rangle
(n^2-(n^{eq})^2)~,
\label{boltzmann}
\end{equation}
where $H$ is the Hubble parameter, $v$ is the `relative 
velocity' of the annihilating particles, $\langle\cdot
\cdot\cdot\rangle$ denotes thermal averaging and 
$\sigma_{eff}$ is the effective cross section defined by
\begin{equation}
\sigma_{eff}=\sum_{i,j}\sigma_{ij}r_ir_j~,
\label{sigmaeff1}
\end{equation}
with $\sigma_{ij}$ being the total cross section for 
particle $i$ to annihilate with particle $j$ averaged 
over initial spin and particle-antiparticle states.
In our case, $\sigma_{eff}$ takes the form 
\begin{equation}
\sigma_{eff}= 
\sigma_{\tilde\chi\tilde\chi}
r_{\tilde\chi}r_{\tilde\chi}+
4\sigma_{\tilde\chi\tilde\tau_2}
r_{\tilde\chi}r_{\tilde\tau_2}+
2(\sigma_{\tilde\tau_2\tilde\tau_2}+
\sigma_{\tilde\tau_2\tilde\tau_2^\ast})
r_{\tilde\tau_2}r_{\tilde\tau_2}~.
\label{sigmaeff2}
\end{equation}
For $r_i$, we use the nonrelativistic approximation
\begin{equation}
r_i(x) = \frac{g_i (1+\Delta_i)^{3/2} e^{-\Delta_i x}}
{g_{eff}}~,
\label{ri}
\end{equation}
\begin{equation}
g_{eff}(x)={\sum_{i}g_i (1+\Delta_i)^{3/2} 
e^{-\Delta_i x}},
~\Delta_i=(m_i-m_{\tilde\chi})/m_{\tilde\chi}~.
\label{geff}
\end{equation}
Here $g_i=2$, 1, 1 ($i=\tilde\chi$, $\tilde\tau_2$, 
$\tilde\tau_2^\ast$) is the number of degrees of 
freedom of the particle species $i$ with mass $m_i$ and 
$x=m_{\tilde\chi}/T$ with $T$ being the photon 
temperature.
 
\par
In Table I, we list all the Feynman graphs included in the 
calculation of the effective cross section. The exchanged 
particles are indicated for each relevant pair of initial and 
final states. The symbols $s(x)$, $t(x)$ and $u(x)$ denote 
tree-graphs in which the particle $x$ is exchanged in 
the s-, t- or u-channel. The symbol $c$ stands for `contact' 
diagrams with all four external legs meeting at a vertex. 
$H$ and $H^\pm$ denote the heaviest neutral and the charged 
higgs bosons, while $e$, $\tilde e_R$, $u$ and $d$ represent 
the first and second generation charged leptons, charged right 
handed sleptons, up- and down-type quarks. The other possible 
reactions $\tilde\tau_2\tilde\tau_2^\ast\rightarrow h[H]A,
~h[H]\gamma,~h[H]Z,~AZ,~H^-W^+~{\rm{or}}~\nu\bar\nu$ 
($\nu$ stands for all three neutrinos) have not been included 
since they are utterly suppressed by small couplings and/or 
heavy masses. Also, the tiny contributions from graphs with $h$ 
and $H$ exchange in the s-channel, in the cases of $u\bar u$, 
$d\bar d$, $e\bar e$ final states, are left out. Some of the 
graphs listed here have not been considered in previous works 
\cite{ellis} with small $\tan\beta$.
\begin{center}
TABLE I. Feynman Diagrams 
\end{center} 
\begin{center}
\begin{tabular}{|c|c|c|} 
\cline{1-1}\cline{2-2}\cline{3-3} 
\multicolumn{1}{|c|}{Initial State} &
\multicolumn{1}{|c|}{Final State} &
\multicolumn{1}{|c|}{Diagrams}\\
\cline{1-1}\cline{2-2}\cline{3-3}
$\tilde\chi\tilde\chi$ & $\tau\bar\tau$ & 
$t(\tilde\tau_{1,2}),~u(\tilde\tau_{1,2})$
\\
& $e\bar e$ & $t(\tilde e_R),~u(\tilde e_R)$
\\ 
\hline  
$\tilde\chi\tilde\tau_2$ & $\tau h,~\tau H,
~\tau Z$ & 
$s(\tau),~t(\tilde\tau_{1,2})$
\\
& $\tau A$ & $s(\tau),~t(\tilde\tau_1)$
\\
& $\tau\gamma$ & 
$s(\tau),~t(\tilde\tau_2)$
\\ 
\hline
$\tilde\tau_2\tilde\tau_2$ & $\tau\tau$ & 
$t(\tilde\chi),~u(\tilde\chi)$
\\ 
\hline
$\tilde\tau_2\tilde\tau_2^\ast$ & 
$~hh,~hH,~HH,~ZZ~$ & $~s(h),~s(H),
~t(\tilde\tau_{1,2}),~u(\tilde\tau_{1,2}),~c~$
\\
& $AA$ & $~s(h),~s(H),~t(\tilde\tau_1),
~u(\tilde\tau_1),~c$
\\   
& $H^+ H^-,~W^+ W^-$ & $s(h),~s(H),~s(\gamma),~s(Z),~c$
\\ 
& $\gamma\gamma,~\gamma Z$ & $t(\tilde\tau_2),
~u(\tilde\tau_2),~c$
\\   
& $t\bar t,~b\bar b$ &
$s(h),~s(H),~s(\gamma),~s(Z)$
\\   
& $\tau\bar\tau$ & $s(h),~s(H),~s(\gamma),~s(Z),
~t(\tilde\chi)$    
\\   
& $u\bar u,~d\bar d,~e \bar e$ & $s(\gamma),~s(Z)$
\\   
\cline{1-1}\cline{2-2}\cline{3-3}                  
\end{tabular}
\end{center}            

\par
The relic abundance of the LSP at the present cosmic time 
can be calculated from the equation \cite{coan,kt} 
\begin{equation}
\Omega_{\tilde\chi}~h^2\approx\frac{1.07 \times 10^9 
~{\rm GeV}^{-1}}{g_*^{1/2}M_{P}~x_F^{-1}~
\hat\sigma_{eff}}
\label{omega}
\end{equation}
with
\begin{equation}
\hat\sigma_{eff}\equiv x_F\int_{x_F}^{\infty}
\langle\sigma_{eff}v\rangle x^{-2}dx~.
\label{sigmaeff3}
\end{equation}
Here $M_P=1.22 \times 10^{19}$ GeV is the Planck scale, 
$g_*\approx 81$ is the effective number of massless 
degrees of freedom at freeze-out \cite{kt} and 
$x_F=m_{\tilde\chi}/T_{F}$, with $T_F$ being the 
freeze-out photon temperature calculated by 
solving iteratively the equation \cite{kt,gkt}
\begin{equation}
x_F = \ln\frac{0.038~g_{eff}(x_F)~M_P~(c+2)c~
m_{\tilde\chi}~\langle \sigma_{eff} v \rangle 
(x_F)}{g_*^{1/2}x_F^{1/2}}~\cdot
\label{xf}
\end{equation}
The constant $c$ is chosen to be equal to $1/2$ \cite{gkt}. 
The freeze-out temperatures which we obtain here are of the 
order of $m_{\tilde\chi}/25$ and, thus, our nonrelativistic 
approximation (see Eq.(\ref{ri})) is justified. Under 
these circumstances, the quantities $\sigma_{ij}v$ are well 
approximated by their Taylor expansion up to second order in 
the `relative velocity', 
\begin{equation}
\sigma_{ij}v=a_{ij}+b_{ij} v^2~.
\label{taylorv}
\end{equation}
The thermally averaged cross sections are then easily 
calculated 
\begin{equation}
\langle \sigma_{ij} v \rangle (x)= 
\frac{x^{3/2}}{2 \sqrt{\pi}} 
\int_{0}^{\infty} d v v^2 (\sigma_{ij} v) e^{-x v^2/4}
=a_{ij}+6 b_{ij}/x~.
\label{average}
\end{equation}
Using Eqs.(\ref{sigmaeff1}), (\ref{sigmaeff2}), 
(\ref{sigmaeff3}) and (\ref{average}), one obtains 
\begin{equation}
\hat\sigma_{eff}=\sum_{(ij)}(\alpha_{(ij)}a_{ij}+
\beta_{(ij)}b_{ij})\equiv
\sum_{(ij)}\hat\sigma_{(ij)}~,
\label{sigmaeff4}
\end{equation}
where we sum over $(ij)=(\tilde\chi\tilde\chi)$, 
$(\tilde\chi\tilde\tau_2)$ and $(\tilde\tau_2
\tilde\tau_2^{(\ast)})$ with 
$a_{\tilde\tau_2\tilde\tau_2^{(\ast)}}=
a_{\tilde\tau_2\tilde\tau_2}+
a_{\tilde\tau_2\tilde\tau_2^{\ast}}$, 
$b_{\tilde\tau_2\tilde\tau_2^{(\ast)}}=
b_{\tilde\tau_2\tilde\tau_2}+
b_{\tilde\tau_2\tilde\tau_2^{\ast}}$ and 
$\alpha_{(ij)}$, $\beta_{(ij)}$ given by
\begin{equation}
\alpha_{(ij)}=c_{(ij)}x_F\int_{x_F}^\infty 
\frac{dx}{x^2}r_i(x)r_j(x)~,
~\beta_{(ij)}=6c_{(ij)}x_F\int_{x_F}^\infty 
\frac{dx}{x^3}r_i(x)r_j(x)~\cdot
\label{alphabeta}
\end{equation}
Here $c_{(ij)}=1$, 4, 2 for $(ij)=
(\tilde\chi\tilde\chi)$, $(\tilde\chi\tilde\tau_2)$ and 
$(\tilde\tau_2\tilde\tau_2^{(\ast)})$.
For $\Delta_{\tilde\tau_2}=0$, $\alpha_{(ij)}=1/4$, 1/2, 
1/8 ($(ij)=(\tilde\chi\tilde\chi)$, 
$(\tilde\chi\tilde\tau_2)$, 
$(\tilde\tau_2\tilde\tau_2^{(\ast)})$), while 
$\beta_{(ij)}=3\alpha_{(ij)}/x_F$.

\subsection{Annihilation cross section}
\label{subsec:annih}

\par
The fact that the LSP ($\tilde\chi$) is an almost 
pure $\tilde B$ implies that the main contribution 
to its annihilation cross section arises from sfermion 
(squark, slepton) exchange in the t- and u-channel 
leading to $f\bar f$ final states ($f$ is a quark 
or lepton). The s-channel diagrams are suppressed 
since the values of $m_{\tilde\chi}$ obtained here 
are always far from $m_Z/2$ and $m_{h}/2$ (see e.g., 
Ref.\cite{drees}). Moreover, diagrams with quarks in 
the final state are suppressed relative to the ones 
with leptons because of the heavier masses of the 
exchanged squarks and the smaller quark hypercharges. 
As mentioned in Sec.\ref{sec:model}, under the 
assumption of unification of the third family Yukawa 
couplings, $m_{\tilde\tau_2}$ is smaller than the 
masses of the other sleptons, hence the production of 
$\tau\bar\tau$ is enhanced relative to the production 
of lighter leptons.

\par
Using the partial wave expansion of Ref.\cite{drees} and 
neglecting the masses of the final state leptons, we 
evaluate the coefficients $a_{\tilde\chi\tilde\chi}$ and 
$b_{\tilde\chi\tilde\chi}$ in Eq.(\ref{taylorv}).
They are found to be
\begin{equation}
a_{\tilde\chi\tilde\chi}=\frac{e^4}{2\pi c_W^4}~ 
s_\theta^2 c_\theta^2 Y_L^2 Y_R^2 
m_{\tilde\chi}^2\left(\frac{1}{\Sigma_2}-
\frac{1}{\Sigma_1}\right)^2,
\label{a11}
\end{equation}
\begin{eqnarray*}    
b_{\tilde\chi\tilde\chi}=\frac{e^4}{12 \pi c_W^4}~ 
\frac{m_{\tilde\chi}^2}{\Sigma_2^4}
\left[(s_\theta^4 Y_L^4+c_\theta^4 Y_R^4)
(m_{\tilde\chi}^4+m_{\tilde\tau_2}^4)+
\frac{s^2_\theta c^2_\theta Y^2_L Y_R^2}{2}
(m_{\tilde\chi}^4+9m_{\tilde\tau_2}^4-
2 m_{\tilde\chi}^2 m_{\tilde\tau_2}^2)\right]
\end{eqnarray*}
\begin{equation}
+2~\frac{e^4}{12\pi c_W^4}~\frac{m_{\tilde\chi}^2 
(m_{\tilde\chi}^4 + m_{\tilde e_R}^4)}{\Sigma^4_e}~,
\label{b11}
\end{equation}
where $Y_{L(R)}=-1/2(-1)$ is the hypercharge of 
$\tau_{L(R)}$, $\Sigma_{1,2}=m_{\tilde\chi}^2+
m_{\tilde\tau_{1,2}}^2$ and $\Sigma_e=
m_{\tilde\chi}^2+m_{\tilde e_R}^2$ with 
$m_{\tilde e_R}$ being the common (see below) mass of 
the right handed sleptons $\tilde e_R$, $\tilde\mu_R$ 
of the two lighter families. Some comments are now in order:

\begin{list}
\setlength{\rightmargin=0cm}{\leftmargin=0cm}

\item[{\bf i.}]
The presence of a nonvanishing coefficient
$a_{\tilde\chi\tilde\chi}$ is due to the large values of 
$\tan\beta$ which lead to an enhancement of the off-diagonal 
terms in the stau mass-squared matrix in Eq.(\ref{stau}). 
Indeed, this coefficient is negligible in the case of small 
$\tilde\tau_L-\tilde\tau_R$ mixing (i.e., for low 
$\tan\beta$) where the   
$\tilde\tau_2$ essentially coincides with $\tilde\tau_R$. 
This is due to the fact that the s-wave contribution, which 
is the only contribution to $a_{\tilde\chi\tilde\chi}~$, 
is suppressed by factors of the final state fermion mass 
as one can show by employing Fermi statistics 
arguments \cite{gold}. For large $\tan\beta$, 
however, this suppression is not complete and 
$a_{\tilde\chi\tilde\chi}$ is proportional to 
$\sin^2\theta$. Despite the fact that 
$a_{\tilde\chi\tilde\chi}$ is smaller than 
$b_{\tilde\chi\tilde\chi}~$, its contribution to 
$\hat\sigma_{eff}$ in Eq.(\ref{sigmaeff4}) 
is of the same order of magnitude as the one of 
$b_{\tilde\chi\tilde\chi}$ which enters in this 
equation divided by a relative factor 
$\stackrel{_{<}}{_{\sim }}x_F/3\sim 8-9$. 

\item[{\bf ii.}]
The main contribution to $b_{\tilde\chi\tilde\chi}$ arises 
from the first term in the bracket in the right hand side 
of Eq.(\ref{b11}). The second term in this bracket is due 
to $\tilde\tau_L-\tilde\tau_R$ mixing.

\item[{\bf iii.}]
The last term in the right hand side of Eq.(\ref{b11}) 
represents the contribution of the two lighter generations. 
Their right handed sleptons are considered degenerate with 
mass $m_{\tilde e_R}$. The off-diagonal elements in the 
slepton mass-squared matrices of the lighter families 
are negligible. The values of $m_{\tilde e_R}$ are bigger 
than $m_{\tilde\tau_2}$ and hence the corresponding 
contributions to $b_{\tilde\chi\tilde\chi}$ are smaller 
than the ones from the $\tilde\tau_2$ exchange. This is a 
major difference from models with low $\tan\beta$, where 
the contributions of all three diagrams with exchange 
of right handed sleptons are similar.

\item[{\bf iv.}]
The contribution to $b_{\tilde\chi\tilde\chi}$ of 
the diagram with a $\tilde\tau_1$ exchange is small 
and, although taken into account in the computation, is 
not displayed in Eq.(\ref{b11}). We find  
that this contribution is suppressed by about $1/6-1/8$ 
compared to the contribution of each of the lightest 
generations. This can be understood by the following 
observation. Despite the fact that the values of the mass 
in the propagator of this diagram, $m_{\tilde\tau_1}$, 
are not much higher than $m_{\tilde e_R}$, its main 
contribution contains a factor $c_\theta^4 Y_L^4$. 

\end{list} 

\subsection{Coannihilation Cross Sections} 
\label{subsec:coan}

\par
The contributions of the various coannihilation processes 
listed in Table I to the coefficients $a_{ij}$ and 
$b_{ij}$ ($ij\neq\tilde\chi\tilde\chi$) in 
Eq.(\ref{taylorv}) are calculated using techniques 
similar to the ones in Ref.\cite{rw}. Leptons and quarks 
(except the $t$-quark) in final states or propagators are 
taken to be massless. On the contrary, the $b$ and $\tau$ 
Yukawa couplings are not ignored since, in our case where 
$\tan\beta$ is large, their influence turns out to be 
very significant. The most important contributions to 
$\hat\sigma_{eff}$ in Eq.(\ref{sigmaeff4}) arise 
from the $a_{ij}$ 's in the case of coannihilation. In 
Table II, we list some of the processes contributing to 
the $a_{ij}$ 's ($ij\neq\tilde\chi\tilde\chi$) 
together with the analytical expressions for their 
contributions.
\begin{center}
TABLE II. Contributions to the Coefficients $a_{ij}$ 
($ij\neq\tilde\chi\tilde\chi$)
\end{center}
\begin{center}
\begin{tabular}{|c|c|}\cline{1-1} \cline{2-2}
\multicolumn{1}{|c|}{Process} &
\multicolumn{1}{|c|}{Contribution to the Coefficient 
$a_{ij}$}\\
\cline{1-1} \cline{2-2}
\hline 
$\tilde\chi\tilde\tau_2\rightarrow\tau h$ & 
$e^2(1-\bar m_h^2)^2\{ 2Y_LY_Rg_{h\tau\tau}
[2s_\theta c_\theta g_h/(m_{\tilde\tau_2}
-\bar m_h^2m_{\tilde\chi})-\cos2\theta\bar g_{h1}/$
\\
& $(\bar m_{\tilde\tau_1}^2+\bar m_{\tilde\chi}
(\bar m_{\tilde\tau_2}-\bar m_h^2))]
+[g_{h\tau\tau}^2+g_h^2/(m_{\tilde\chi}\bar m_h^2-
m_{\tilde\tau_2})^2]
(s_\theta^2 Y_L^2+c_\theta^2 Y_R^2)$ 
\\
& $+\bar g_{h1}^2(c_\theta^2Y_L^2+s_\theta^2 Y_R^2)/
(\bar m_{\tilde\tau_1}^2+\bar m_{\tilde\chi}
(\bar m_{\tilde\tau_2}-\bar m_h^2))^2
-2s_\theta c_\theta (Y_L^2-Y_R^2)$
\\
& $\bar g_h \bar g_{h1}/
(\bar m_{\tilde\tau_2}-\bar m_{\tilde\chi}\bar m_h^2)
(\bar m_{\tilde\tau_1}^2+\bar m_{\tilde\chi}
(\bar m_{\tilde\tau_2}-\bar m_h^2))\}
/32\pi c_W^2 m_{\tilde\tau_2}
(m_{\tilde\tau_2}+m_{\tilde\chi})$
\\ 
\hline
$\tilde\chi\tilde\tau_2\rightarrow\tau\gamma$ &
$e^4(s_\theta^2Y_L^2+c_\theta^2Y_R^2)/
16\pi c_W^2 m_{\tilde\tau_2}
(m_{\tilde\chi}+m_{\tilde\tau_2})$
\\ 
\hline
$\tilde\chi\tilde\tau_2\rightarrow\tau Z$ &
$e^2(1-\bar m_Z^2)\{\bar m_{\tilde\tau_2}
(1-\bar m_Z^2)^3[g_{\tilde\tau_2 \tilde\tau_2 Z}^2
(s_\theta^2Y_L^2+c_\theta^2 Y_R^2)
/(\bar m_Z^2\bar m_{\tilde\chi}-
\bar m_{\tilde\tau_2})^2$
\\
& $+g_{\tilde\tau_1\tilde\tau_2Z}^2 
(c_\theta^2Y_L^2+ s_\theta^2Y_R^2)/
(\bar m_{\tilde\tau_1}^2+\bar m_{\tilde\chi}
(\bar m_{\tilde\tau_2}-\bar m_Z^2))^2$
\\
& $-2g_{\tilde\tau_1\tilde\tau_2Z} 
g_{\tilde\tau_2\tilde\tau_2Z}
s_\theta c_\theta (Y_L^2-Y_R^2)/ 
(\bar m_{\tilde\tau_2}-\bar m_Z^2\bar m_{\tilde\chi})
(\bar m_{\tilde\tau_1}^2+\bar m_{\tilde\chi}
(\bar m_{\tilde\tau_2}-\bar m_Z^2))]$
\\
& $-2g_Z(\bar m_Z^2-1)^2
[g_{\tilde\tau_2\tilde\tau_2Z}
(L_\tau s_\theta^2Y_L^2 +R_\tau c_\theta^2Y_R^2)
/(\bar m_{\tilde\tau_2}
-\bar m_Z^2\bar m_{\tilde\chi})$
\\
& $-g_{\tilde\tau_1\tilde\tau_2Z}s_\theta c_\theta  
(L_\tau Y_L^2-R_\tau Y_R^2)/
(\bar m_{\tilde\tau_1}^2+\bar m_{\tilde\chi}
(\bar m_{\tilde\tau_2}-\bar m_Z^2))]$
\\
& $+g_Z^2(L_\tau^2s_\theta^2Y_L^2+R_\tau^2c_\theta^2Y_R^2)
(1+\bar m_Z^2-2\bar m_Z^4)(1+\hat m_{\tilde\chi})\}
/32\pi c_W^2 m_Z^2$
\\ 
\hline 
$\tilde\tau_2\tilde\tau_2\rightarrow\tau\tau$ & 
$e^4(s_\theta^4Y_L^4+c_\theta^4Y_R^4)m_{\tilde\chi}^2
/\pi c_W^4 \Sigma_2^2$ 
\\ 
\hline  
$\tilde\tau_2\tilde\tau_2^\ast\rightarrow
\gamma\gamma$ & 
$e^4/8\pi m_{\tilde\tau_2}^2$
\\ 
\hline
$\tilde\tau_2\tilde\tau_2^\ast\rightarrow\gamma Z$ & 
$-e^2g_{\tilde\tau_2\tilde\tau_2 Z}^2(\hat m_Z^2-4)
/16 \pi m_{\tilde\tau_2}^2$
\\ 
\hline
$\tilde\tau_2\tilde\tau_2^\ast\rightarrow ZZ$ & 
$(1-\hat m_Z^2)^{1/2}
\{[(g_h^2g_{hZZ}^2P_1/(\hat m_h^2-4)
+12g_hg_{hZZ}g_{\tilde\tau_2\tilde\tau_2Z}^2
m_Z^2\hat m_Z^2)/(\hat m_h^2-4)$
\\
& $-4g_hg_{hZZ}g_{\tilde\tau_1\tilde\tau_2Z}^2
m_{\tilde\tau_2}^2(P_4-\hat m_{\tilde\tau_1}^2P_1)
/(1+\hat m_{\tilde\tau_1}^2-\hat m_Z^2)
(\hat m_h^2-4)+\left(h\leftrightarrow H\right)]$
\\
& $+g_{\tilde\tau_2\tilde\tau_2Z}^4m_Z^4P_3/
(\hat m_Z^2-2)^2+2g_hg_{hZZ}g_Hg_{HZZ}P_1/
(\hat m_h^2-4)(\hat m_H^2-4)$
\\
& $-8g_{\tilde\tau_2\tilde\tau_2Z}^2
g_{\tilde\tau_1\tilde\tau_2Z}^2
m_Z^4[P_5-3\hat m_{\tilde\tau_1}^2(\hat m_Z^2-2)]/
(1+\hat m_{\tilde\tau_1}^2-\hat m_Z^2)(\hat m_Z^2-2)$
\\
& $+4m_{\tilde\tau_2}^4
g_{\tilde\tau_1\tilde\tau_2Z}^4
[\hat m_{\tilde\tau_1}^4P_1+(1-\hat m_Z^2)^2P_2
-2\hat m_{\tilde\tau_1}^2P_4]
/(1+\hat m_{\tilde\tau_1}^2-\hat m_Z^2)^2 \}/$
\\
& $64\pi m_Z^4 m_{\tilde\tau_2}^2$ 
\\ 
\hline 
$\tilde\tau_2\tilde\tau_2^\ast\rightarrow W^+ W^-$ & 
$(1-\hat m_W^2)^{1/2}
(4-4\hat m_W^2+3\hat m_W^4)[g_hg_{hW^+ W^-}/
(\hat m_h^2-4)$
\\
& $+g_Hg_{HW^+ W^-}/(\hat m_H^2-4)
+g_{\tilde\tau_2\tilde\tau_2 W^+ W^-}
m_{\tilde\tau_2}^2]^2/32\pi m_W^4 m_{\tilde\tau_2}^2$
\\ 
\hline
$\tilde\tau_2\tilde\tau_2^\ast\rightarrow t\bar t$ & 
$3(1-\hat m_t^2)^{3/2}[ g_h g_{htt}/(\hat m_h^2-4)
+g_Hg_{Htt}/(\hat m_H^2-4)]^2/4\pi m_{\tilde\tau_2}^4$ 
\\ 
\hline 
\end{tabular}
\end{center}
Here hat (or bar) over a quantity indicates that this quantity 
is measured in units of $m_{\tilde\tau_2}$ (or 
$m_{\tilde\chi}+m_{\tilde\tau_2}$) and the $g$ 's will be 
defined shortly. Also, $L_\tau=1-2 s_W^2$, $R_\tau=-2s_W^2$ 
and 
$$
P_{1(2)}=3\hat m_Z^4-(+)4\hat m_Z^2+4~,  
~P_3=3\hat m_Z^4-8\hat m_Z^2+8~,
$$
\begin{equation}
P_4=3\hat m_Z^6-3\hat m_Z^4-4\hat m_Z^2+4~, 
~P_5=3\hat m_Z^4-5\hat m_Z^2+2~.
\label{pi}
\end{equation}

\par
The contribution of the process 
$\tilde\chi\tilde\tau_2\rightarrow\tau H$ (or $\tau A$) 
to the coefficient $a_{\tilde\chi\tilde\tau_2}$ is 
obtained from the expression for $\tilde\chi\tilde\tau_2
\rightarrow\tau h$ in Table II by replacing $h$ by $H$ 
(or $A$ and $\cos 2\theta$ by 1). For the contribution to 
$a_{\tilde\tau_2\tilde\tau_2^\ast}$ of each of the five 
processes with two higgses in the final state (see Table I), 
a general formula can be given:
$$
a_{\tilde\tau_2\tilde\tau_2^\ast\rightarrow H_pH_q}=
(\frac{1}{2})\frac{1}
{128 \pi m_{\tilde\tau_2}^6}
(4-(\hat m_{H_p}-\hat m_{H_q})^2)^{1/2}
(4-(\hat m_{H_p}+\hat m_{H_q})^2)^{1/2}
$$
\begin{equation}
\left(\frac{\lambda_h}{4-\hat m_h^2}
+\frac{\lambda_H}{4-\hat m_H^2}+
\frac{4\lambda_1}{\hat m_{H_p}^2+\hat m_{H_q}^2
-2\hat m_{\tilde\tau_1}^2-2}
+\frac{4\lambda_2}{\hat m_{H_p}^2+\hat m_{H_q}^2-4}
-\lambda_c m_{\tilde\tau_2}^2\right)^2,
\label{ahiggs}
\end{equation}
where the $H_p$, $H_q$ stand for $h$, $H$, $A$, $H^+$, 
$H^-$, the factor $1/2$ enters only for identical particles 
in the final state and $\lambda_h$, $\lambda_H$, 
$\lambda_1$, $\lambda_2$, $\lambda_c$ correspond to the 
diagrams $s(h)$, $s(H)$, $t(\tilde\tau_{1,2})$ 
(or $u(\tilde\tau_{1,2})$), $c$ in Table I and are shown 
in the Table III.
\begin{center}
 Table III. The $\lambda$ Symbols 
\end{center} 
\begin{center}
\begin{tabular}{|c|c|c|c|c|c|}
\cline{1-1}\cline{2-2}\cline{3-3}\cline{4-4}\cline{5-5}
\cline{6-6}
\multicolumn{1}{|c|}{ Process} &
\multicolumn{1}{|c|}{ $\lambda_h$ }&
\multicolumn{1}{|c|}{ $\lambda_H$ }&
\multicolumn{1}{|c|}{ $\lambda_1$ }&
\multicolumn{1}{|c|}{ $\lambda_2$ }&
\multicolumn{1}{|c|}{ $\lambda_c$ }\\
\hline 
$\tilde\tau_2\tilde\tau_2^\ast\rightarrow hh$ & 
$g_hg_{hhh}$ & 
$g_Hg_{hhH}$ & 
$g_{h1}^2$ & 
$g_h^2$ &
$g_{\tilde\tau_2\tilde\tau_2 hh}$ 
\\ 
\hline
$\tilde\tau_2\tilde\tau_2^\ast\rightarrow hH$ & 
$g_hg_{hhH}$ &
$g_Hg_{hHH}$ & 
$g_{h1}g_{H1}$ & 
$g_hg_H$ &
$g_{\tilde\tau_2\tilde\tau_2 hH}$
\\ 
\hline
$\tilde\tau_2\tilde\tau_2^\ast\rightarrow HH$ & 
$g_hg_{hHH}$ & 
$g_Hg_{HHH}$ & 
$g_{H1}^2$ &
$g_H^2$ &
$g_{\tilde\tau_2\tilde\tau_2 HH}$ 
\\ 
\hline  
$\tilde\tau_2\tilde\tau_2^\ast\rightarrow AA$ & 
$g_hg_{hAA}$ & 
$g_Hg_{HAA}$ & 
$-g_{A1}^2$ & 
$0$ &
$g_{\tilde\tau_2\tilde\tau_2 AA}$  
\\ 
\hline  
$\tilde\tau_2\tilde\tau_2^\ast\rightarrow H^+ H^-$ & 
$g_hg_{hH^+ H^-}$ & 
$g_Hg_{HH^+ H^-}$ & 
$0$ & 
$0$ &
$g_{\tilde\tau_2\tilde\tau_2 H^+ H^-}$    
\\ 
\hline 
\end{tabular}
\end{center}

\par
The $g$ 's in Tables II and III correspond to vertices 
with the particles entering indicated as subscripts.
The simplest ones are (for Feynman rules, see e.g., 
Ref.\cite{hk} with $\mu\rightarrow-\mu$)
\begin{equation}
g_{\tilde\tau_1\tilde\tau_2Z}=g_Z(-s_\theta c_\theta)~,
~g_{\tilde\tau_2\tilde\tau_2Z}=g_Z(s_\theta^2-2s_W^2)~,
~g_{\tilde\tau_2\tilde\tau_2 W^+ W^-}=g^2 s_\theta^2/2~,
\label{g}
\end{equation}
where $g_Z=g/2 c_W$ with $g$ being the $SU(2)_L$ gauge 
coupling constant. Note that $g_{A}\equiv 
g_{\tilde\tau_2\tilde\tau_2 A}=0$. 
The more complicated $g$ 's are arranged 
in the Table IV.
\begin{center}
{Table IV. $g$ Symbols} 
\end{center} 
\begin{center}
\begin{tabular}{|c|c|}
\cline{1-1}\cline{2-2}
\multicolumn{1}{|c|}{$g$ Symbol} &
\multicolumn{1}{|c|}{Expression}\\
\hline  
$g_{h[H]1}\left(\equiv 
g_{\tilde\tau_1\tilde\tau_2 h[H]}\right)$ & 
$g_Z m_Z \sin[-\cos](\alpha+\beta)
(L_\tau+R_\tau)s_\theta c_\theta$
\\
& $+gm_\tau\cos 2\theta(A_\tau\sin[-\cos]\alpha
-\mu\cos[\sin]\alpha)/2m_W \cos\beta$
\\ 
\hline
$g_{h[H]}\left(\equiv 
g_{\tilde\tau_2\tilde\tau_2 h[H]}\right)$ & 
$-g_Z m_Z\sin[-\cos](\alpha+\beta)
(L_\tau s_\theta^2-R_\tau c_\theta^2)-
(gm_\tau/m_W\cos\beta)$
\\
& $\{-m_\tau\sin[-\cos]\alpha
-s_\theta c_\theta(A_\tau\sin[-\cos]\alpha-
\mu\cos[\sin]\alpha)\}$
\\ 
\hline  
$g_{A1}\left(\equiv 
g_{\tilde\tau_1\tilde\tau_2 A}\right)$ & 
$gm_\tau(A_\tau\tan\beta-\mu)/2m_W$
\\ 
\hline
$g_{\tilde\tau_2\tilde\tau_2 hh[HH]}$ & 
$ -[+]g_Z^2 \cos 2\alpha
(L_\tau s_\theta^2-R_\tau c_\theta^2)
-g^2 (\sin[\cos]\alpha/\cos\beta)^2m_\tau^2/2m_W^2$
\\
\hline 
$g_{\tilde\tau_2\tilde\tau_2 hH}$ & 
$g^2 \sin 2\alpha(-L_\tau/2c_W^2
+m_\tau^2/2m_W^2\cos^2\beta)s_\theta^2/2$
\\
& $+g^2\sin 2\alpha(-\tan^2\theta_W
+m_\tau^2/2m_W^2\cos^2\beta)c_\theta^2/2$
\\ 
\hline
$g_{\tilde\tau_2\tilde\tau_2 AA}$ & 
$-g_Z^2 \cos2\beta 
(L_\tau s_\theta^2-R_\tau c_\theta^2)
-g^2 \tan^2\beta (m_\tau/m_W)^2/2$
\\ 
\hline 
$g_{\tilde\tau_2\tilde\tau_2 H^+ H^-}$ & 
$g^2\cos 2\beta ((1-L_\tau/2c_W^2)s_\theta^2
-\tan^2\theta_Wc_\theta^2)/2$  
\\
& $-g^2\tan^2\beta (m_\tau/m_W)^2 c_\theta^2/2$
\\ 
\hline
$g_{hhh[HHH]}$ & 
$-3g_Z m_Z \sin[\cos](\alpha+\beta)\cos 2\alpha$ 
\\ 
\hline 
$g_{hHH[hhH]}$ & 
$g_Z m_Z \{\sin[\cos](\alpha+\beta)\cos 2\alpha
+2\cos[-\sin](\alpha+\beta)\sin 2\alpha \}$ 
\\ 
\hline 
$g_{h[H]AA}$ 
& $-g_Z m_Z\sin [-\cos](\alpha+\beta)\cos 2\beta$
\\ 
\hline 
$g_{h[H]H^+ H^-}$ & 
$-g\{ m_W\sin[\cos](\beta-\alpha) 
+m_Z\sin[-\cos](\alpha+\beta)\cos 2\beta/2c_W\}$ 
\\ 
\hline  
$g_{h[H]ZZ}$ & 
$g m_Z\sin[\cos](\beta-\alpha)/c_W$
\\ 
\hline 
$g_{h[H]W^+ W^-}$ & 
$g m_W \sin[\cos](\beta-\alpha)$ 
\\ 
\hline 
$g_{h[H]tt}$ & 
$-g(\cos[\sin]\alpha/\sin\beta)(m_t/2m_W)$
\\ 
\hline
$g_{h[H]\tau\tau}$ & 
$g(\sin[-\cos]\alpha/\cos\beta)(m_\tau/2m_W)$
\\ 
\hline
$g_{A\tau\tau}$ & 
$-g\tan\beta(m_\tau/2m_W)$
\\ 
\hline
\end{tabular}
\end{center}
Here we have defined
\begin{equation}
\tan 2\alpha=\tan 2\beta~(m_A^2+m_Z^2)/(m_A^2-m_Z^2)~,
~-\pi/2\leq\alpha\leq 0~.
\label{alpha}
\end{equation}

\par
We do not show explicitly the small contributions to 
$a_{\tilde\tau_2\tilde\tau_2^\ast}$ of the processes 
with $b\bar b$ and $\tau\bar\tau$ in the final state. 
They are, however, taken into account in the computation. 
The contributions to 
$a_{\tilde\tau_2\tilde\tau_2^\ast}$ of the processes 
with $u\bar u$, $d\bar d$ and $e\bar e$ in the final 
state vanish (these processes contribute only to $b$ 's). 
Also, the coefficients $b_{ij}$ 
($ij\neq\tilde\chi\tilde\chi$), although included in 
the calculation, are not displayed since their contribution 
to $\hat\sigma_{eff}$ is, in general, negligible. Note 
that many of the couplings and terms listed above have not
been included in previous calculations \cite{ellis} with 
small $\tan\beta$. Some comments are now in order:

\begin{list}
\setlength{\rightmargin=0cm}{\leftmargin=0cm}

\item[{\bf i.}]
All five processes for the coannihilation of 
$\tilde\chi$ with $\tilde\tau_2$ listed 
in Table I give more or less comparable contributions 
to the coefficient $a_{\tilde\chi\tilde\tau_2}$ (the 
leading contribution comes, in general, from 
$\tilde\chi\tilde\tau_2\rightarrow\tau h$). The 
relative contribution of $b_{\tilde\chi\tilde\tau_2}$ 
to $\hat\sigma_{(\tilde\chi\tilde\tau_2)}$ 
in Eq.(\ref{sigmaeff4}) turns out to be essentially 
independent of $m_A$ (95 GeV$\leq m_A\leq$ 
220 GeV). This contribution varies from about $5\%$ 
to about $8\%$ as $\Delta_{\tilde\tau_2}$ 
increases from 0 to 0.1 (this is the relevant range of 
$\Delta_{\tilde\tau_2}$ as we shall see).

\item[{\bf ii.}]
The major contributions to 
$a_{\tilde\tau_2\tilde\tau_2^{(\ast)}}$ come from the 
processes $\tilde\tau_2\tilde\tau_2^{\ast}\rightarrow
hh,~t\bar t$ and $\tilde\tau_2\tilde\tau_2\rightarrow
\tau\tau$. However, many of the other relevant processes 
in Table I (like $\tilde\tau_2\tilde\tau_2^{\ast}
\rightarrow ZZ$, $\gamma\gamma$, $HH$, $AA$, 
$H^+H^-$, $\gamma Z$) have, in general, important 
contributions which cannot be neglected ($\tilde\tau_2
\tilde\tau_2^{\ast}\rightarrow ZZ$, for large 
$\Delta_{\tilde\tau_2}$ 's and $m_A$ 's, gives a 
major contribution). Also, the 
reaction $\tilde\tau_2\tilde\tau_2^{\ast}\rightarrow 
hH$ ($W^+W^-$) is enhanced for low values of $m_A$ (and 
$\Delta_{\tilde\tau_2}$). The relative contribution of 
$b_{\tilde\tau_2\tilde\tau_2^{(\ast)}}$ to
$\hat\sigma_{(\tilde\tau_2\tilde\tau_2^{(\ast)})}$, 
which can be either positive or negative, is less than about 
$1\%$ for all relevant values of parameters.

\item[{\bf iii.}]
For $\Delta_{\tilde\tau_2}=0$, the contribution of the 
$\tilde\chi$ annihilation to $\hat\sigma_{eff}$ is 
very small ($\approx 0.4\%$). The corresponding 
contributions of 
$\hat\sigma_{(\tilde\chi\tilde\tau_2)}$ and 
$\hat\sigma_{(\tilde\tau_2\tilde\tau_2^{(\ast)})}$ 
span the ranges $27-24\%$ and $73-76\%$ respectively as 
$m_A$ varies from 95 to 220 GeV. For 
$\Delta_{\tilde\tau_2}=0.1$, however, the annihilation 
of $\tilde\chi$ 's becomes very significant accounting for 
about $33-31\%$ of $\hat\sigma_{eff}$. The most important 
contribution ($\approx 58\%$ of $\hat\sigma_{eff}$), 
in this case, comes from the coannihilation of $\tilde\chi$ 
with $\tilde\tau_2$, whereas $\tilde\tau_2$ coannihilation 
with $\tilde\tau_2$ or $\tilde\tau_2^{\ast}$ accounts 
for about $9-11\%$ of $\hat\sigma_{eff}$. We see that, 
although $\tilde\chi$ annihilation is negligible 
for small $\Delta_{\tilde\tau_2}$ 's, it is strongly 
enhanced at higher values of $\Delta_{\tilde\tau_2}$. 
This is due to the fact that the abundance of 
$\tilde\tau_2$ 's decreases relative to the one of 
$\tilde\chi$ 's as $\Delta_{\tilde\tau_2}$ increases.

\item[{\bf iv.}]
For $\Delta_{\tilde\tau_2}=0$, the contributions of
$b_{\tilde\chi\tilde\tau_2}$ and $b_{\tilde\tau_2
\tilde\tau_2^{(\ast)}}$ to $\hat\sigma_{eff}$
cancel each other partially and, thus, an accurate 
result (error $\approx 0.5\%$) can be obtained by 
ignoring these $b$ 's. For $\Delta_{\tilde\tau_2}=0.1$, 
however, the contribution of $b_{\tilde\chi\tilde\tau_2}$ 
dominates strongly over the one of 
$b_{\tilde\tau_2\tilde\tau_2^{(\ast)}}$ and gives 
$\approx 4-5\%$ of $\hat\sigma_{eff}$. Consequently,
our results can be reproduced with an accuracy better than 
$\approx 5\%$ by using, for coannihilation, just the $a$ 's. 
Their analytical expressions have been given earlier in this 
Section. On the contrary, the 
$b_{\tilde\chi\tilde\chi}$ cannot be ignored since its 
contribution to $\hat\sigma_{(\tilde\chi\tilde\chi)}$ 
can be as high as $80\%$ and the annihilation of 
$\tilde\chi$ 's is very significant at higher 
$\Delta_{\tilde\tau_2}$ 's.

\end{list}

\subsection{Results on $\Omega_{LSP}~h^2$}
\label{subsec:result}

\par
\begin{figure}
\epsfig{figure=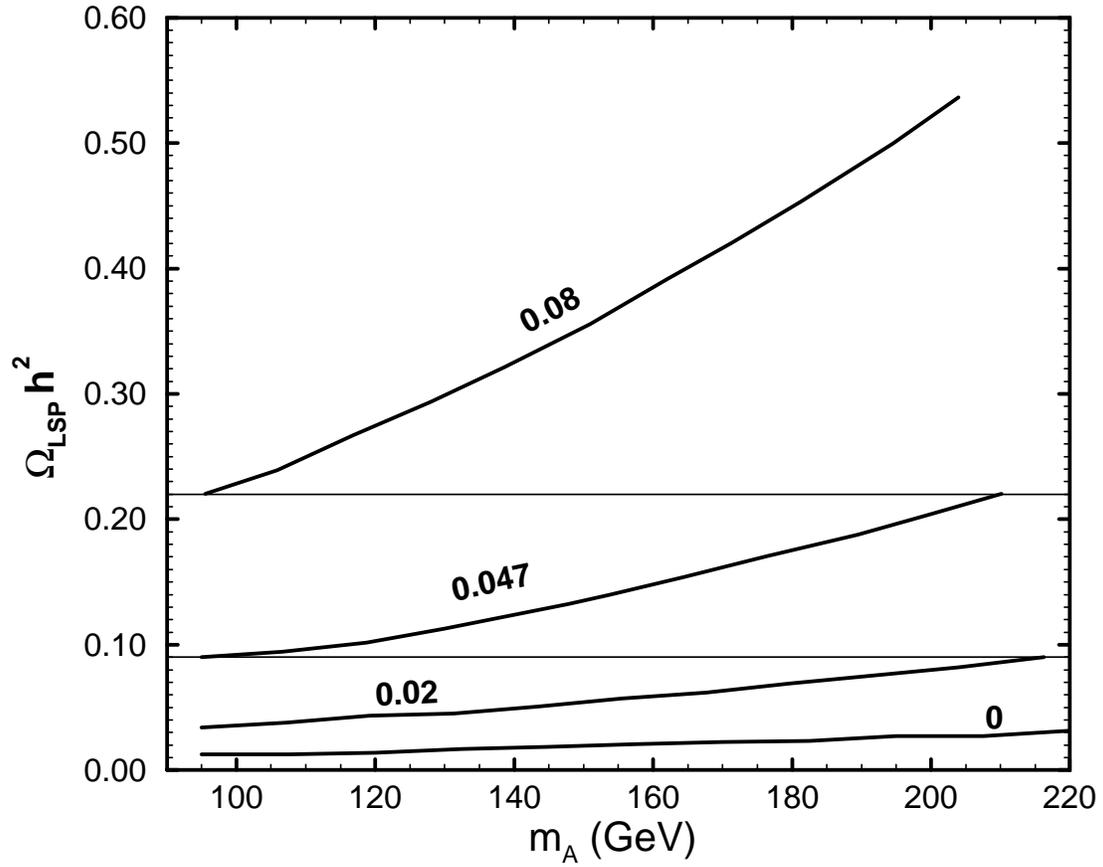,height=4.62in,angle=0}
\medskip
\caption{The LSP abundance $\Omega_{LSP}~h^2$ as a 
function of $m_A$ for $\Delta_{\tilde\tau_2}=0$, 
0.02, 0.047 and 0.08 as indicated. The limiting lines
at $\Omega_{LSP}~h^2=0.09$ and 0.22 are also included.
\label{lsp}}
\end{figure}

We can now proceed to the evaluation of 
$\Omega_{LSP}~h^2$. The top quark mass $m_t(m_t)$ is 
again fixed at 166 GeV. For given values of 
$\Delta_{\tilde\tau_2}$ and $m_A$, all the particle 
physics parameters of the model are determined (see 
Sec.\ref{sec:model}). The freeze-out parameter $x_F$ 
can then be found by solving Eq.(\ref{xf}) and 
$\hat\sigma_{eff}$ is evaluated from 
Eq.(\ref{sigmaeff4}). The LSP relic abundance 
$\Omega_{\tilde\chi}~h^2$ is estimated using 
Eq.(\ref{omega}) and is depicted in Fig.\ref{lsp} as
function of $m_A$ for $\Delta_{\tilde\tau_2}$=0, 
0.02, 0.047 and 0.08. Remember that the curves 
on this plot, which correspond to specific values of 
$\Delta_{\tilde\tau_2}$, terminate at the appropriate 
upper limit on $m_A$ (derived from the restriction 
$M_{1/2}\leq 800~{\rm{GeV}}$). This limit decreases 
as $\Delta_{\tilde\tau_2}$ increases. 

\par
Requiring $\Omega_{\tilde\chi}~h^2$ to be 
confined in the cosmologically allowed range $0.09-0.22$, 
one finds that $\Delta_{\tilde\tau_2}$ is restricted 
between $\approx$ 0.02 and 0.08. Note that the upper 
limit on $\Delta_{\tilde\tau_2}$ does not depend on 
our restriction on $M_{1/2}$. On the contrary, the 
lower limit on $\Delta_{\tilde\tau_2}$ is somewhat 
dependent on the particular choice one makes for this 
restriction. This deserves further study which would 
require going beyond the simplifying assumption of a 
common supersymmetry threshold $M_S$. It should be 
pointed out that this lower bound on 
$\Delta_{\tilde\tau_2}$ is anyway evaded if there 
exist additional contributions to the cold dark matter 
of the universe from particle species other than 
$\tilde\chi$.  

\par
\begin{figure}
\epsfig{figure=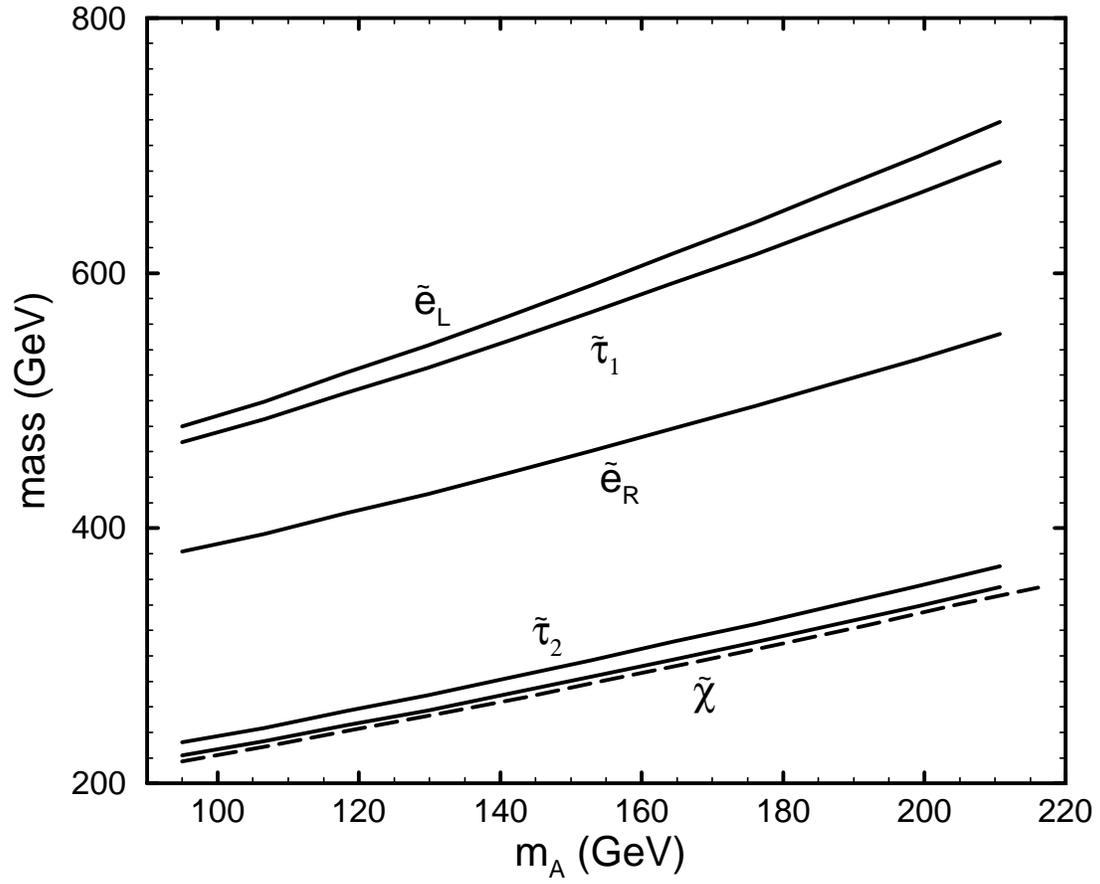,height=4.7in,angle=0}
\medskip
\caption{The relevant part of the sparticle spectrum as a 
function of $m_A$ for $\Delta_{\tilde\tau_2}= 0.047$. 
The LSP mass, for $\Delta_{\tilde\tau_2}= 0.02$, is 
also included (dashed line).
\label{spectrum}}
\end{figure}
\begin{figure}
\epsfig{figure=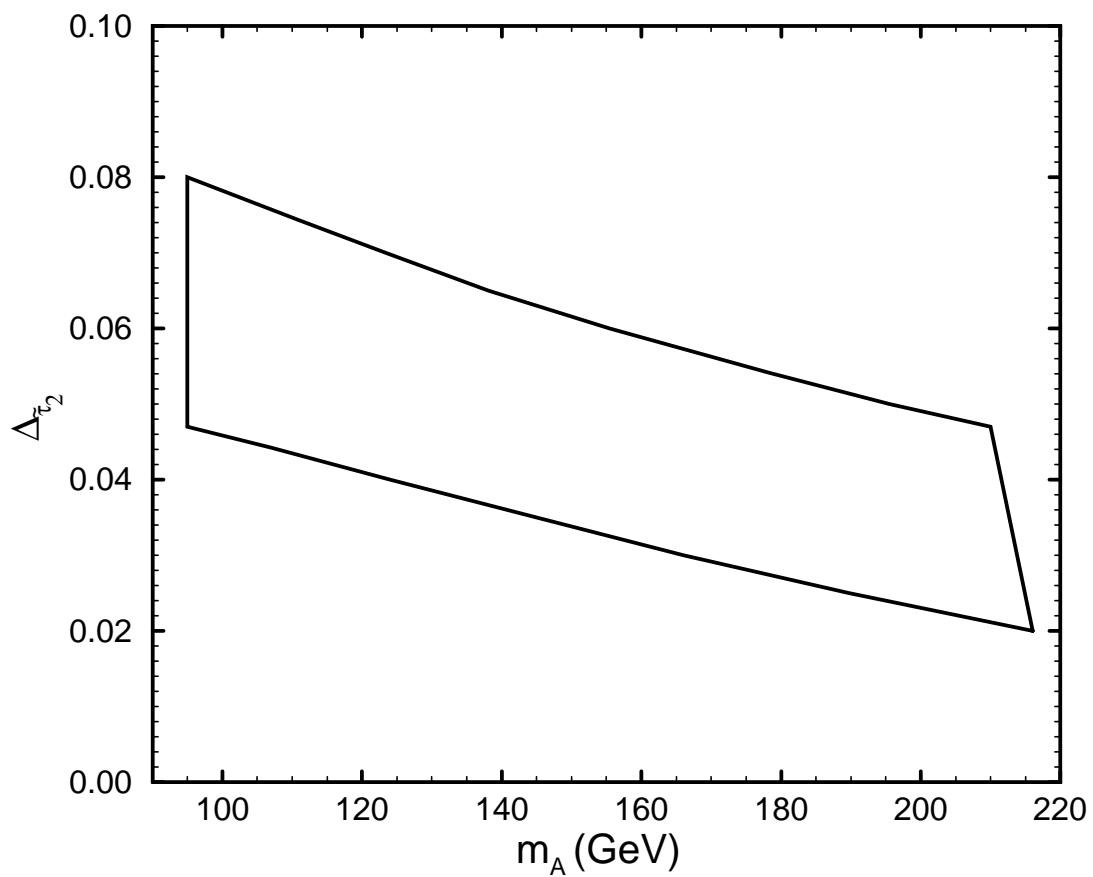,height=4.645in,angle=0}
\medskip
\caption{The cosmologically allowed region in the 
$m_A-\Delta_{\tilde\tau_2}$ plane, where $\Omega_{LSP}~h^2$ 
lies in the range $0.09-0.22$. We also take $m_A\geq 95$ GeV 
and $M_{1/2}\leq 800$ GeV.
\label{delta}}
\end{figure}

Fig.\ref{delta} shows the cosmologically allowed region 
in the $m_A-\Delta_{\tilde\tau_2}$ plane 
obtained from the above considerations.  We see that 
$m_A$ can vary only between about 95 and 216 GeV. The 
lower (upper) boundary of this region corresponds to 
$\Omega_{\tilde\chi}~h^2=0.09$ (0.22). The left 
boundary corresponds to $M_{1/2}=800~{\rm{GeV}}$ 
($0.09\leq\Omega_{\tilde\chi}~h^2\leq 0.22$). Along 
this line, $m_{\tilde\chi}$ is essentially constant and 
acquires its maximal allowed value $\approx 354~{\rm{GeV}}$ 
(see Fig.\ref{spectrum}). The minimal value of the LSP 
mass is obtained at the lower left corner of this allowed 
region, where $\Delta_{\tilde\tau_2}\approx 0.047$, and is 
equal to about 222 GeV (see Fig.\ref{spectrum}). We, thus, 
see that the LSP mass ranges between $\approx$ 222 and 354 
GeV. The $\tilde\tau_2$ mass is bounded between about 232 
and 369 GeV which makes $\tilde\tau_2$ a phenomenologically 
interesting charged sparticle. The upper (lower) bound 
corresponds to the upper right (lower left) 
corner of the region in Fig.\ref{delta}. Actually, the 
whole sparticle mass spectrum is strongly restricted by 
our considerations. Note, however, that the 
upper bounds on the sparticle masses depend on our choice 
for the maximal allowed $M_{1/2}$. This requires a  
detailed study with inclusion of one-loop and supersymmetry 
threshold effects which may not be negligible for higher 
$M_{1/2}$ 's.

\par
\begin{figure}
\epsfig{figure=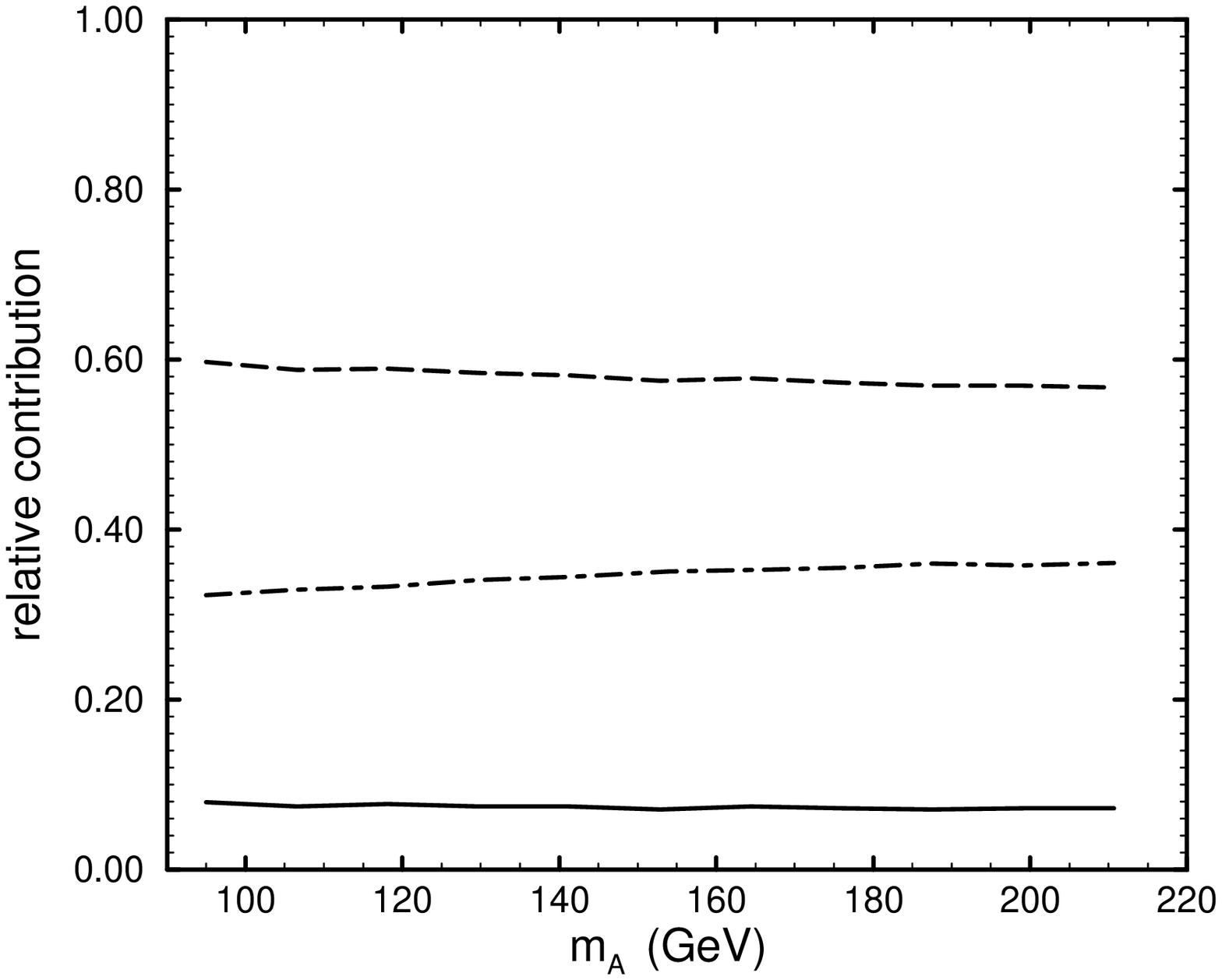,height=4.65in,angle=0}
\medskip
\caption{The relative contributions 
$\hat\sigma_{(\tilde\chi\tilde\chi)}/\hat\sigma_{eff}$ 
(solid line), $\hat\sigma_{(\tilde\chi\tilde\tau_2)}
/\hat\sigma_{eff}$ (dashed line) and 
$\hat\sigma_{(\tilde\tau_2\tilde\tau_2^{(\ast)})}/
\hat\sigma_{eff}$ (dot-dashed line) of the three inclusive 
(co)annihilation reactions to $\hat\sigma_{eff}$ as 
functions of $m_A$ for $\Delta_{\tilde\tau_2}=0.047$. 
\label{sigma}}
\end{figure}

The relative contributions $\hat\sigma_{(ij)}/
\hat\sigma_{eff}$ ($(ij)=(\tilde\chi\tilde\chi)$, 
$(\tilde\chi\tilde\tau_2)$, 
$(\tilde\tau_2\tilde\tau_2^{(\ast)})$) of the three 
inclusive (co)annihilation reactions to $\hat\sigma_{eff}$ 
are given in Fig.\ref{sigma} as functions of $m_A$ for the 
`central' value of $\Delta_{\tilde\tau_2}=0.047$. The 
allowed range of $m_A$ is $95-211$ GeV in this case.       

\section{Conclusions}
\label{sec:conclusion}

\par
We considered the MSSM with gauge and Yukawa 
coupling unification employing radiative electroweak 
symmetry breaking with universal boundary conditions from 
gravity-mediated supersymmetry breaking. We calculated the 
relic density of the LSP (an almost pure bino). 
Coannihilation of the LSP with the NLSP (the lightest 
stau) is crucial for reducing its relic density to an 
acceptable level. Compatibility with 
the mixed or the pure cold (with a nonzero 
cosmological constant) dark matter scenarios for 
structure formation in the universe requires 
that the lightest stau mass is about $2-8\%$ larger 
than the bino mass. This combined with the fact that the 
LSP mass is restricted to be greater than about 222 GeV 
allows the lightest stau mass to be as low as 232 GeV.

\acknowledgements{
We thank B. Ananthanarayan, C. Boehm, M. Drees, N. Fornengo, 
S. Khalil, A. B. Lahanas, K. Olive and N. D. Vlachos for 
discussions. One of us (C. P.) thanks P. Porfyriadis for 
computational help and the Greek State Scholarship 
Institution (I. K. Y) for support. This work 
was supported by the European Union under TMR contract 
No. ERBFMRX--CT96--0090 and the Greek Government research 
grant PENED/95 K.A.1795.}

\def\ijmp#1#2#3{{ Int. Jour. Mod. Phys. }{\bf #1~}(19#2)~#3}
\def\pl#1#2#3{{ Phys. Lett. }{\bf B#1~}(19#2)~#3}
\def\zp#1#2#3{{ Z. Phys. }{\bf C#1~}(19#2)~#3}
\def\prl#1#2#3{{ Phys. Rev. Lett. }{\bf #1~}(19#2)~#3}
\def\rmp#1#2#3{{ Rev. Mod. Phys. }{\bf #1~}(19#2)~#3}
\def\prep#1#2#3{{ Phys. Rep. }{\bf #1~}(19#2)~#3}
\def\pr#1#2#3{{ Phys. Rev. }{\bf D#1~}(19#2)~#3}
\def\np#1#2#3{{ Nucl. Phys. }{\bf B#1~}(19#2)~#3}
\def\mpl#1#2#3{{ Mod. Phys. Lett. }{\bf #1~}(19#2)~#3}
\def\arnps#1#2#3{{ Annu. Rev. Nucl. Part. Sci. }{\bf
#1~}(19#2)~#3}
\def\sjnp#1#2#3{{ Sov. J. Nucl. Phys. }{\bf #1~}(19#2)~#3}
\def\jetp#1#2#3{{ JETP Lett. }{\bf #1~}(19#2)~#3}
\def\app#1#2#3{{ Acta Phys. Polon. }{\bf #1~}(19#2)~#3}
\def\rnc#1#2#3{{ Riv. Nuovo Cim. }{\bf #1~}(19#2)~#3}
\def\ap#1#2#3{{ Ann. Phys. }{\bf #1~}(19#2)~#3}
\def\ptp#1#2#3{{ Prog. Theor. Phys. }{\bf #1~}(19#2)~#3}
\def\plb#1#2#3{{ Phys. Lett. }{\bf#1B~}(19#2)~#3}
\def\apjl#1#2#3{{Astrophys. J. Lett. }{\bf #1~}(19#2)~#3}
\def\n#1#2#3{{ Nature }{\bf #1~}(19#2)~#3}
\def\apj#1#2#3{{Astrophys. Journal }{\bf #1~}(19#2)~#3}
\def\anj#1#2#3{{Astron. J. }{\bf #1~}(19#2)~#3}
\def\mnras#1#2#3{{MNRAS }{\bf #1~}(19#2)~#3}
\def\grg#1#2#3{{Gen. Rel. Grav. }{\bf #1~}(19#2)~#3}
\def\s#1#2#3{{Science }{\bf #1~}(19#2)~#3}
\def\baas#1#2#3{{Bull. Am. Astron. Soc. }{\bf #1~}(19#2)~#3}
\def\ibid#1#2#3{{ibid. }{\bf #1~}(19#2)~#3}

\newpage

\pagestyle{empty}

\begin{references}
\bibitem{mb} This problem arises in the simplest 
realization of the $SO(10)$ model with universal boundary 
conditions and Yukawa coupling unification, where the 
electroweak higgses exclusively belong to a $SO(10)$ 
10-plet. However, in a complete $SO(10)$ model of this 
variety, which correctly incorporates fermion masses and 
mixings, the electroweak higgses will receive smaller 
contributions from other representations too. Under some 
circumstances (see, e. g., G. Lazarides and Q. Shafi, 
\np{350}{91}{179}), these contributions can affect only 
the down quark and lepton masses leaving, in particular, 
the top quark mass unaltered. The details of the scheme 
can then be arranged so that $m_b(m_Z)$ becomes 
experimentally acceptable, while $m_{\tau}$ and 
$\tan\beta$ are minimally affected. Alternatively, 
small GUT threshold corrections to gauge (or Yukawa) 
coupling unification can also help to reduce $m_b$ in 
specific models without abandoning universal boundary 
conditions. An example with a small negative GUT 
correction to $\alpha_s$ and acceptable value of 
$m_b$ can be found in T. Bla\v{z}ek and S. Raby, 
\pr{59}{99}{095002}. It is interesting that such a 
correction to $\alpha_s$ is anyway needed for gauge 
coupling unification \cite{pierce}. Our LSP relic 
density calculation is not affected in any essential 
way by these modifications of the simplest scheme.
\end{references}
\end{document}